# A chip-scale oscillation-mode optomechanical inertial sensor near the thermodynamical limits

*Yongjun Huang\*, Jaime Gonzalo Flor Flores\*, Ying Li, Wenting Wang, Di Wang, Noam Goldberg, Jiangjun Zheng, Mingbin Yu, Ming Lu, Michael Kutzer, Daniel Rogers, Dim-Lee Kwong, Layne Churchill, and Chee Wei Wong\**

Dr. Y. Huang

School of Information and Communication Engineering, University of Electronic Science and Technology of China, Chengdu, 611731, China

E-mail: yongjunh@uestc.edu.cn

Dr. Y. Huang, J. G. Flor Flores, Dr. Wenting Wang, Prof. C. W. Wong

Fang Lu Mesoscopic Optics and Quantum Electronics Laboratory, University of California, Los Angeles, CA 90095, USA

E-mail: jflorflores@ucla.edu; cheewei.wong@ucla.edu

Dr. Y. Huang, Dr. Y. Li, D. Wang, N. Goldberg, Dr. J. Zheng, Prof. C. W. Wong

Optical Nanostructures Laboratory, Columbia University, New York, NY 10027, USA

Prof. M. Yu, Prof. D.-L. Kwong

Institute of Microelectronics, A*STAR, Singapore 117865

Prof. M. Lu

Brookhaven National Laboratory, Upton, NY 11973, USA

Dr. M. Kutzer, D. Rogers, L. Churchill

Johns Hopkins Applied Physics Laboratory, Laurel, MD 20723, USA

**Abstract:** High-precision inertial sensing and gravity sensing are in navigation, oil exploration, and earthquake prediction. Modern navigation systems integrate the global positioning system (GPS) with an inertial navigation system (INS), which complement each other for correct attitude and velocity determination. The core of the INS is the inertial measurement unit which integrates accelerometers and gyroscopes used to measure the forces and angular rate in the vehicular inertial reference frame. With the help of the gyroscopes and by integration of the acceleration to compute velocity and distance, precision and compact accelerometers with sufficient accuracy in the minute-timescale can provide small-error location determination. Solid-state implementations, through coherent readout, can provide a platform with significantly more massive resonators for acceleration detection. In contrast to prior accelerometers using piezoelectric or electronic capacitance readout techniques, optical readout provides narrow-linewidth high-sensitivity laser detection along with low-noise resonant optomechanical transduction near the thermodynamical limits. Here an optomechanical inertial sensor with 8.2-μ$g$/Hz$^{1/2}$ velocity random walk (VRW) at acquisition rate of 100 Hz and 50.9 μ$g$ bias instability is demonstrated, suitable for consumer and industrial grade applications, e.g., inertial navigation, inclination sensing, platform stabilization, and/or wearable device motion detection. Driven into optomechanical sustained-oscillation, the slot photonic crystal cavity provides radio-frequency readout of the optically-driven transduction with enhanced 625-µ$g$/Hz sensitivity. Measuring the optomechanically-stiffened oscillation shift, instead of the optical transmission shift, provides a 220× VRW enhancement over pre-oscillation mode detection due to the strong optomechanical transduction. Supported by theory, this inertial sensor operates 2.56× above the thermodynamical limit at small integration times, with 43-dB dynamic range, in a solid-state room-temperature readout architecture.

## 1. Introduction

Laser-cooled atoms and quantum interferometry has provided new advances in precision measurements of fundamental constants [1], time [2], forces [3], gravity [4,5], gravitational red-shift [6], and inertial sensing [7], In parallel, recent advances in radiation-pressure driven cavity optomechanics [8,9] have provided new frontiers for laser cooling of mesoscopic systems [10–12], chip-scale stable RF sources [13–15], phonon lasers [16], induced-transparency through multi-mode interferences [17,18], chaos generation and transfer [19,20], and explorations into potential quantum transductions of microwave, spin, and optical qubits [21,22]. In cavity optomechanics, the optical cavity and mechanical resonator are co-designed to achieve large optomechanical coupling and transduction [9,23]. This optomechanical transduction is described by the optomechanical coupling coefficient $g_{om}/2\pi$ – a shift of the optical frequency $\omega_o$ per unit motional displacement $x$, or equivalently d$\omega_o$/d$x$. With carefully-tuned high-quality optical factor ($Q_o$) and tight sub-wavelength confinement of selected optical cavity modes, the induced radiation pressure or optical gradient force can modify or drive the motion of micromechanical resonators [8,9]. As a consequence, a change in optical gradient force also modifies the effective micromechanical resonance frequency. The coupled optical and mechanical degrees-of-freedom [24] allow detections of nanomechanical motion [25,26] towards force sensing [27], radio wave detection [28], and magnetic susceptibility measurements [29].

In the past several years, high-performance micro-accelerometers based on cavity optomechanics have been remarkably advanced for acceleration sensing at higher frequencies above 100 Hz for vibration or shock measurements. Such examples include nanobeam cavities at 10 µ$g$/Hz$^{1/2}$ (1 $g$ = 9.81 m/s$^2$) resolution (sensing bandwidth 5 to 25 kHz) with lock-in detection [30], cantilevered whispering gallery mode (WGM) spherical silica cavities at 4.5 µ$g$/Hz$^{1/2}$ resolution (sensing bandwidth 10 to 100 Hz) [31], and bulk fiber-optic cavities at 100 n$g$/Hz1/2 resolution (sensing bandwidth 1 to 10 kHz) [32,33]. Optomechanical accelerometers with external piezoelectric stages and calibration have also been tested with outdoor operation [34-37].

Alternatively, we measure the optomechanically-stiffened oscillation shift instead, which translate into a mechanical frequency shift. This technique generates advantages such as a larger signal-to-noise ratio to suppress the flicker noise and better oscillation frequency shift tracking precision that contributes to improved resolution. In this paper, an oscillation-mode optomechanical inertial sensor with 8.2-µ$g$/Hz$^{1/2}$ velocity random walk (VRW) at an acquisition rate of 100 Hz and 50.9 µ$g$ bias instability is demonstrated. Driven by an optically-induced gradient force, our tight sub-wavelength confinement in the photonic crystal cavity [38] allows for a large optomechanical transduction coefficient ($g_{om}/2\pi$) [23] while allowing concurrently a large motional mass for precision sensing. Driven into optomechanical sustained-oscillation and measuring the RF shift instead of the optical resonance, the slot photonic crystal cavity provides RF readout of the optically-driven transduction with enhanced 625-µg/Hz sensitivity, and a 220× VRW enhancement over pre-oscillation mode detection; all of this due to the strong optomechanical transduction and readout in the optical domain. Our optomechanical inertial sensor operates 2.56× above the thermodynamical limit at small integration times and a non-flat noise density, with a 43-dB dynamic range in a solid-state room-temperature vacuum ($\approx 10^{-6}$ torr) architecture, and is supported by our theory and numerical modeling.

## 2. Results

### 2.1. Design of a laser-driven RF-readout optomechanical inertial sensor

The mechanism of our inertial sensor is based on a laser-driven silicon photonic crystal cavity as shown in Figure 1a, with one side of the cavity on a stationary beam and the other side of the cavity on a large motional mass. When an acceleration is applied to the motional mass, the slot width *s* is perturbed, shifting the optical resonance frequency and changing the detuning from the drive laser frequency. This perturbs the intracavity optical energy and hence the optomechanical stiffening [detailed below in Equation (1)]. The resulting shifted RF resonance frequency is subsequently read out with high precision, close to the thermodynamical limits.

The optomechanical inertial sensor consists of a slot-type photonic crystal cavity, as illustrated in Figure 1b, nanofabricated in a silicon-on-insulator substrate with a 250 nm thick silicon layer (see Methods). The cavity is formed by shifted and perturbed lattice holes [23,38] designed to form the localized optical resonant modes as well as an optomechanical gradient force transduction. The slot cavity features a slot width of *s* = 80 nm, holes with a radii of *r* = 150 nm and lattice constant of $a_{pc}$ = 470 nm, with lattice perturbations of 5 nm (red), 10 nm (green) and 15 nm (blue) to form the localized cavity modes on a line-defect waveguide with width of 1.2 ×$\sqrt{3}a_{pc}$. The cavity mode volume is subwavelength at $0.051(\lambda/n)^3$. The large ($\approx 120$ µm × 150 µm) proof mass of $m_x$ = 5.6 nanogram for acceleration detection has four (1 µm × 50 µm) compliant support beams providing a fundamental mechanical resonance of $\Omega_m \approx 2\pi \times 71.3$ kHz and a combined stiffness of $k \approx 1.12$ N/m. One side of the slot cavity is attached to this mass and named as section *i* in Figure 1a. The other side (*ii*) of the slot cavity is cantilevered to the substrate and has the same *x*-length as the proof mass to reduce asymmetric residual stress *z*-bow between the two sections (*i* and *ii*), in order to preserve the localized optical resonance mode.

### 2.2. Basic characterizations of the optomechanical inertial sensor at zero-acceleration

Initially the performance of the designed optomechanical inertial sensor with large proof mass is demonstrated and characterized at a zero-acceleration condition. Probed by a dimpled tapered fiber anchored to side *ii* of the optomechanical cavity, Figure 1c shows the measured optical transmission spectra under different driving powers. Under a low-driving power (22 µW), the loaded cavity quality factor $Q_o$ is measured at $\approx 4{,}350$. With increasing driving power, thermal nonlinearity broadens the cavity resonances into asymmetric line-shapes. With optical pump powers greater than 188 µW, the optomechanical cavity is driven into self-induced regenerative oscillations[13] – with significantly narrower mechanical oscillation linewidths and signature spectral fluctuations.

The mechanical motion of the proof mass in the optomechanical inertial sensor (section *i* in Figure 1a), excited by the optical gradient force through strong optomechanical transduction, is recorded by using a fast optical detector and a spectrum analyzer. Figure 2a shows an example RF power spectrum, with the optical transmission simultaneously monitored using dual photodetectors, illustrating the fundamental mechanical resonance of 71.3 kHz for the proof mass. The left inset of Figure 2a shows the Lorentzian curve fitting, which shows a vacuum ($10^{-6}$ torr) mechanical quality factor $Q_m$ at $\approx$ 1383, for the fundamental mode; this is largely bounded by the large proof mass and anchor losses. The proof mass displacement is described by $x(\Omega) = (\Omega_m^2 - \Omega^2 + i\Omega\Omega_m/Q_m)^{-1} a(\Omega) = \chi(\Omega) a(\Omega)$ when subjected to an external low-frequency acceleration and thermal noise: $a(\Omega) = a_o(\Omega) + a_{dc} + a_{th}$ in the low frequency regime ($\Omega \ll \Omega_m$), $\chi(\Omega) = 1/\Omega_m^2$. The right *y*-axis of Figure 2a and its inset also show the corresponding power spectral density in displacement noise units, reaching a displacement noise floor of $\approx$ 0.2 fm/Hz$^{1/2}$.

In the slot-type optomechanical cavity, resonant enhancement of the pump-laser optical gradient force yields strong backaction, and due to the deeply sub-wavelength confinement, the optomechanical stiffening and optical-RF resonance spectra of the cavity are strongly dependent on the slot width *s*.[23] In the presence of an (+*y* direction) acceleration, side *i* of the membrane of the slot cavity shifts with displacement $x_s$, decreasing the slot width *s*. Consequently, the optomechanical stiffening increases with a resulting increasing RF resonant frequency $\Omega'_m$ described by:

$$\Omega'_m = \sqrt{\Omega_m^2 + \left(\frac{2|\hat{a}|^2 g_{om}^2}{\Delta^2 \omega_c m_x}\right) \Delta_o} = \sqrt{\Omega_m^2 + \left(\frac{2|\hat{a}|^2 g_{om}^2}{((\omega_l - \omega_c + g_{om} x_s)^2 + (\Gamma/2)^2)\omega_c m_x}\right)(\omega_l - \omega_c + g_{om} x_s)} \quad (1)$$

where $\Omega'_m$ ($\Omega_m$) is the shifted (unperturbed) RF resonance frequency, $|\hat{a}|^2$ the averaged intracavity photon energy, $g_{om}$ the optomechanical coupling rate, $m_x$ the effective mass, and $\Gamma = 1/\tau$ and $\omega_c$ the optical cavity decay rate and optical resonance frequency respectively.

From the ($x_s$ – to – $\Omega'_m$) transfer function in Equation 1, we also note the strong dependence of the mechanical power spectra on the optical cavity resonance, intracavity energy, laser detuning $\Delta_o|_{xs=0} = \omega_l - \omega_0$ and $\Delta^2|_{xs=0} = \Delta^2_o + (\Gamma/2)^2$, which enhances the minimal displacement and acceleration sensitivity detection. In addition, we can experimentally monitor the mechanical power spectra under different laser detuning when there is no external acceleration acting on the optomechanical inertial sensor. Figure 2b illustrates the measured 2D RF power spectra with a swept pump wavelength, for the test case of a static (non-acceleration, $x_s = 0$) frame. For the current case, the inertial sensor's optical resonance is 1552.83 nm at a 22 µW incident power with a 50 aJ optical intracavity energy (yielding an intracavity photon number of $\approx$ 425) at zero optical detuning. The modeled mechanical frequency for different laser detuning (using Equation 1) is also superimposed on the 2D measurement. From the measurement-modeling correspondence, we obtain the optomechanical coupling rate to be $g_{om}/2\pi$ = 37.1 GHz/nm, with a vacuum optomechanical coupling rate $g^*/2\pi$ = 180 kHz. We observe that the RF resonance has a strong optical resonance dependence in the detuning range of $\Delta_o = \pm$ 0.2 nm (correlated with the slot width *s*), with a largest mechanical resonance frequency shift $\Delta\Omega_m/2\pi$ of 450 Hz over an externally-driven 70 pm wavelength detuning. We note that the measured $g_{om}/2\pi$ is smaller than theoretical prediction due to elevation asymmetries of the released masses (sections *i* and *ii* in Figure 1a), lowering the optomechanical transduction from design values. Also, with increased injected optical power, the RF shift range is enhanced due to the stronger optical gradient force as illustrated in Figure 2b and 2c, resulting in larger slope and better sensitivity when subjected an external acceleration.

### 2.3. Optomechanical inertial sensor operation in pre-oscillation mode

To measure the RF shift $\Delta\Omega_m$ under low-frequency motion, we fixed the driving wavelength and the driving power (with laser instabilities less than ± 166 fm and ± 0.0017-dB in 10-minutes), and mounted the device and vacuum setup onto a high-torque Aerotech ADRS-200 rotary stage. This is shown in Figure 2d. With optimized digital control and drive filter parameters, the rotary stage provides a noise-equivalent input low-frequency ($\approx$ 20 Hz) acceleration precision down to sub-$\mu g$ (in the +*y* axis direction; smaller slot width). Figure 3a shows the measured backaction RF transduction in the pre-oscillation mode at 40 $\mu$W drive power with $\Omega_m/2\pi \approx 71.5$ kHz in this measurement. Examining the Lorentzian RF line-shapes over the first 40 spectra (each waveform corresponds to an increase in the applied acceleration up to 154.8 m*g*) and the theoretical modeling, the RF frequency is observed to increase by 56.5 Hz on average over the measurement range, equivalent to a sensitivity *S* of 2.62±0.31 m*g*/Hz (or a scale factor of 381.5±45 Hz/*g*).

The minimum detectable low-frequency acceleration for the optomechanical inertial sensor has the well-known definition as $a_{min}$ = $S \cdot \delta\Omega_{rms}/2\pi$,[39,40] where $\delta\Omega_{rms}/2\pi$ is the RMS frequency variation in a specific data acquisition rate, corresponding to the integration time.] $\delta\Omega_{rms}/2\pi$ is determined to be 6.9 Hz. This corresponds to a minimum acceleration noise floor, or the resolution of our optomechanical inertial sensor, of $\approx$ 1.8±0.21 m*g*/Hz$^{1/2}$, in pre-oscillation mode. Figure 3e illustrates the measured bias instability and VRW noise of $\approx$ 5.23 m*g* and 1.8 m*g*/Hz$^{1/2}$ respectively, which are the two very important parameters indicating the minimum achievable low-frequency acceleration reading limited by flicker noise and the total noise density [31,33]. Theoretically, in pre-oscillation mode, the thermal noise frequency fluctuation is given by [41,42]: $\delta\Omega_{th}$ = $[(k_BT/E_C)\cdot(\Omega_m\Delta f/Q_m)]^{1/2}$, from the measured power spectra density and with $\Delta f$ the acquisition rate corresponding to the integration time, and $E_C$ = $m_x\Omega_m^2<x_c^2>$ the maximum energy in the oscillator and $<x_c>$ the constant mean square amplitude. With a signal-to-noise ratio (SNR = 10log($E_C/k_BT$)) of $\approx$ 18 dB and $Q_m$ of 1383, we obtain $\delta\Omega_{th}/2\pi$ as 9.04 Hz in 0.01 second integration time. Correspondingly, with the sensitivity of 1.1 m*g*/Hz for the optimized optical detuning, this gives a theoretically-estimated acceleration resolution $R \approx$ 995 $\mu g$/Hz$^{1/}$, in pre-oscillation mode. The theoretical thermodynamical acceleration noise floor with optimized optical detuning closed to the optical resonance is also presented. The excess noises shown in the measurements are mainly from system noise such as the laser frequency noise, parasitic displacement noise, device thermal fluctuation and thermal expansion noise among others. We note that only the first 40 RF spectra (up to 154.8 m*g*) are used in the analysis to be well within the linear dynamic range (<0.05% standard deviation). We also measured other inertial sensor devices in our setup with similar geometries, including an inertial sensor mounted 180° opposite from Figure 3a (i.e. acceleration in the -*y* axis direction); the mechanical resonance frequency shift is observed in opposite direction under the same acceleration ranges.

## 2.4. Optomechanical inertial sensor operation in oscillation mode

Next we drove the 5.6 nanogram optomechanical inertial sensor into oscillation mode as shown in Figure 3b, 3c and 3d, with driving powers above 188 μW. From the nonlinear optomechanical transduction, high-order harmonics (up to the 30th harmonic) of the fundamental resonance are observed above threshold,[9] with two example spectra shown in the top insets of Figure 3b. In this device, the fundamental oscillation frequency is 85.3 kHz as seen in Figure 3d. Note the ≈ 70-dB intensity peak-to-noise floor when in the self-induced regenerative oscillation regime, compared to ≈ 18 dB in the resonant (pre-oscillation) mode. Figure 3c illustrates the tracked RF shift under 6.53 m$g$ to 26.34 m$g$ low-frequency accelerations, with ≈ 200 μW intracavity driving power. We note the sizably larger RF shift at ≈ 31.7 Hz, compared to the pre-oscillation regime, even when measured over ≈ 11× smaller low-frequency acceleration ranges as compared to Figure 3a. The inset of Figure 3c illustrates the spectra comparison (left to right) under 6.53, 14.68 and 26.34 m$g$ respectively. To characterize the oscillator, we also measured the frequency instability of our optomechanical inertial sensor. This shows an Allan deviation of $3.8 \times 10^{-6}$ fractional frequency instability at 10 ms integration. Oscillation frequency shift dependence shows a detection sensitivity $S$ of 625±4.6 μ$g$/Hz (or equivalently, a scale factor of 1.6±0.012 kHz/$g$). The improved sensitivity is due to the larger slope as shown in Figure 2b and 2c, resulting from the higher drive power ($|\hat{a}|^2$).

The obtained minimum acceleration noise floor in oscillation mode, based on the measured oscillator sensitivity and long-term frequency instability, is determined to be ≈ 8.2±0.07 μ$g$/Hz$^{1/2}$ in the acquisition rate of 100 Hz, and from low integration times of 10 ms to 100 ms. We emphasize that this is ≈ 220× better than the pre-oscillation mode. Figure 3e also shows the determined bias instability of ≈ 50.9 μ$g$ in oscillation mode. In addition to the ± 4.6 μ$g$/Hz sensitivity regression estimate, the long-time (1-hour timescale) standard deviation of the frequency shift fluctuations $\delta\Omega/2\pi$ is also analyzed to be 9.78 Hz. It should be noticed that VRW is mostly related to thermal effects, and it provides a measure of these as seen by the system. At the same time bias instability is one of the most important parameters when constant acceleration is measured for inertial navigation. In oscillation mode, we also note that the large peak-to-noise floor ratio allows a minimal distinguishable oscillation frequency shift (between two overlapping Lorentzian lineshapes) $\delta\Omega_{rms}$ much smaller than in pre-oscillation.

## 3. Discussion

Even in pre-oscillation mode, we emphasize that the detectable 3.05 Hz RF shift is equivalent to a 467 fm wavelength perturbation in the optical cavity resonance. Such 3-Hz RF precision detection requirement is much easier than a 467-fm optical frequency shift because the optical setup would require homodyne detection – extra components essentially. This makes our RF readout optomechanical inertial sensor more suitable for compact and low-cost commercial application. For maximal sensitivity to the external low-frequency acceleration, we can further tune the drive and readout laser wavelength to the slot-cavity resonance. The cavity slot direction is aligned perpendicular to the rotation radius, allowing the imparted centrifugal force to be in the ±$y$-direction onto the proof mass (the alignment error is negligible and, even for a ± 5° error, contributes only a ± 0.38% fixed offset in the imparted force measurement). Furthermore, we note that the estimated instabilities of the driving wavelength (± 166 fm) over 10-minutes contribute to only a ± 5% uncertainty in the acceleration resolution; for long-term measurements this uncertainty can be readily improved with more stable drive lasers. We also note that the coupled intracavity power is monitored to account for any coupling drifts or power fluctuations in the rotation measurements, and a polarization-maintaining tapered fiber serves as the input/output coupling fiber through vacuum chamber to avoid any polarization shifts during the imparted rotation and acceleration.

Without optomechanical stiffening, the inertial sensor resolution is solely represented by the mechanical domain parameters and described by thermal Brownian noise. With the large mass oscillator and high driving powers, the thermal Brownian noise can be largely reduced [41,42] and the measured frequency instability and acceleration sensitivity can be strongly improved, resulting in better resolution. The optical noise bound arising from quantum backaction noise $a_{BA}$ [33] is estimated to be $\approx$ 184 n$g$/Hz$^{1/2}$ for our oscillator design and implementation parameters. We note that, with this RF transduction readout approach instead of solely the optical intensity transmission readout, optical shot noise and photodetector shot noise do not contribute significantly to the fundamental limit in our readout scheme, allowing our measured acceleration noise floor to be close to the thermal noise limit.

To support the inertial sensor demonstration, we determine the linear dynamic range experimentally from the minimum detectable acceleration (8.2 µ$g$) to $\approx$ 170 m$g$, as 43 dB. This differs from a theoretical design estimate of $\approx$ 44.8 due to fabrication variations and the nonlinear dependence of the optical resonance on the slot width $s$. In addition, to further improve the sensitivity and resolution, self-reference noise cancellation [43] or thermal stabilization through multiply-resonant optomechanical oscillators [44] can be pursued. The remaining deviations of the modeled sensitivity from the measurements likely arise from a residual $\approx$ 25% over-estimate of the optical resonance shift per displacement in the finite-difference time-domain simulations.

## 4. Conclusion

We have demonstrated an optomechanical inertial sensor driven into the oscillation mode, for the first time, for enhanced noise-equivalent acceleration measurements. In pre-oscillation mode, the optomechanical drive and readout show a sensitivity of 2.62 m$g$/Hz. Driven into self-regenerative optomechanical oscillation modes, the oscillator demonstrates an enhanced acceleration sensitivity of 625 µ$g$/Hz and a noise floor down to $\approx$ 8.2 µ$g$/Hz$^{1/2}$ with 100 Hz data acquisition rate, close to the thermal noise limit ($\approx$ 3.2 µ$g$/Hz$^{1/2}$) at small integration times. The mesoscopic room-temperature implementation with RF optomechanical transduction and readout provides a platform towards low-noise precision sensing such as inertial navigation, inclination sensing, and platform stabilization.

**Supporting Information**

Supporting Information is available from the Wiley Online Library or from the author.

**Acknowledgements**

We acknowledge discussions with S.-W. Huang, J. Lim, J. F. McMillan, J. Wu, S. Bhave, P.-C. Hsieh, Z. Xie, W. Wang, and P. Hamilton. The authors acknowledge support from the National Science Foundation (SDC-1520952), Johns Hopkins Applied

Physics Laboratory, and the Department of Defense. Research carried out in part at the Center for Functional Nanomaterials, Brookhaven National Laboratory, which is supported by the U.S. Department of Energy, Office of Basic Energy Sciences, under contract number DE-AC02-98CH10886. Y. H. acknowledges the supports from National Natural Science Foundation of China (61701082), National Postdoctoral Program for Innovative Talents (BX201700043), and Sichuan Provincial Science and Technology Planning Program of China (2019YFG0120). Y. Huang, J. G. Flor Flores, and Y. Li contributed equally to this work.

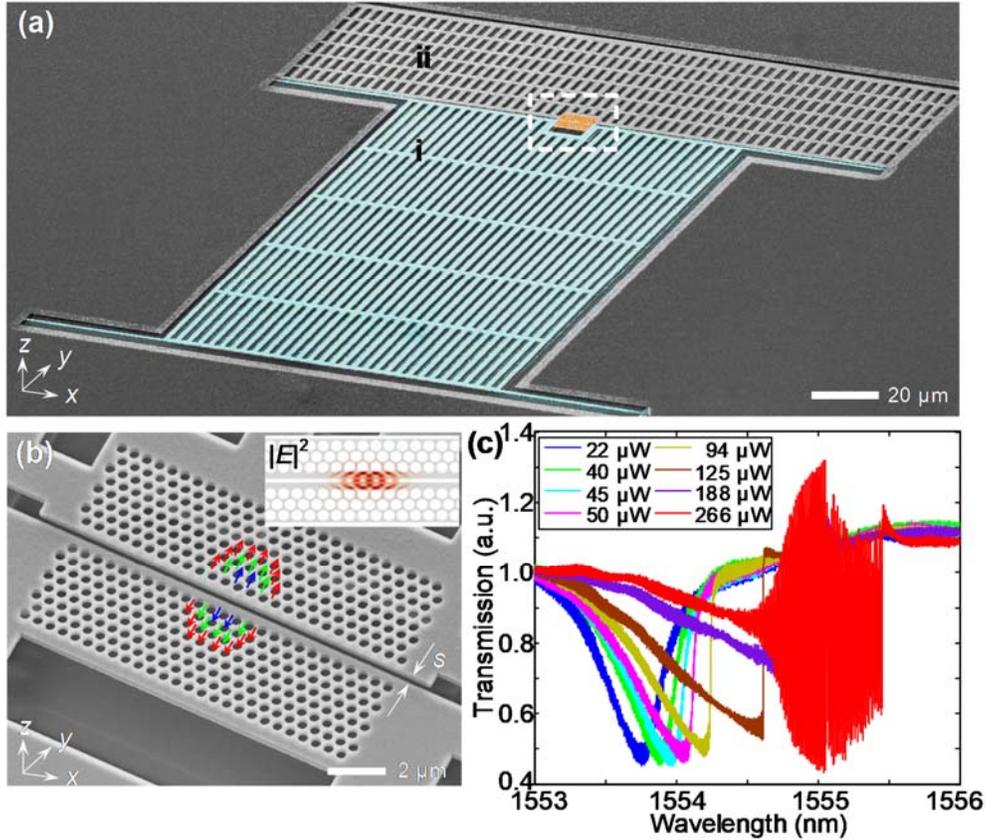

**Figure 1.** A chip-sale optomechanical inertial sensor. a) Scanning electron micrographs (SEMs) of inertial sensor nanofabricated in silicon-on-insulator (250 nm device layer) with a 5.6 nanogram proof mass and co-designed slot-type photonic crystal cavity with a slot width *s* of 80 nm. The fundamental mechanical resonance is designed around 60 to 85 kHz, above most of the ambient, acoustic and seismic vibrational noise frequencies. Section *i* (in cyan) is the moving mass and section *ii* (in grey) is the stationary section. Scale bar: 20 μm. The orange region is the slot-type photonic crystal cavity, with the dashed white box expanded in panel b. b) Colored arrows denote the 5 nm (red), 10 nm (green) and 15 nm (blue) photonic crystal lattice perturbations to form the slot-localized resonance modes with $0.051(\lambda/n)^3$ mode volumes. Scale bar: 2 μm. Inset: computed $|E|^2$-distribution for the designed nanocavity. c) Optical transmission spectra of the inertial sensor, with increasing drive and readout powers (22 μW to 266 μW) including observations of self-induced regenerative oscillations with the signature spectral fluctuations (in red).

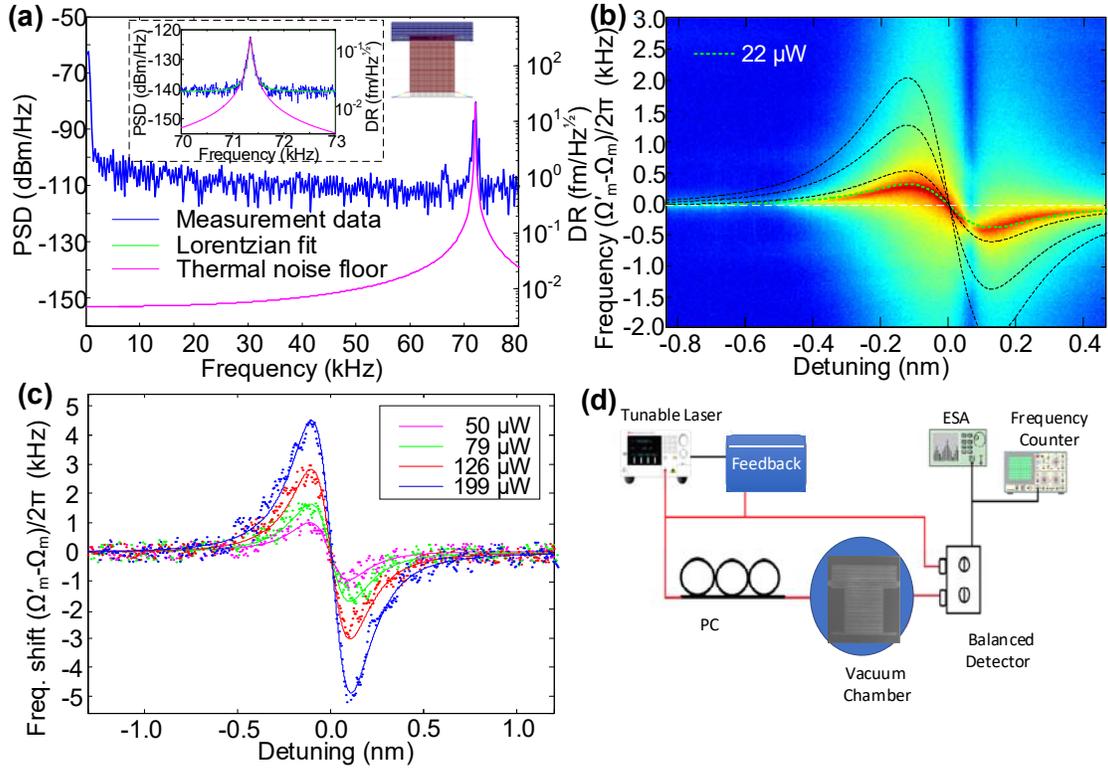

**Figure 2.** Inertial sensor transduction and enhanced sensitivity in the RF domain. a) Optical gradient force backaction transduction into the RF power spectral density for the first fundamental mechanical mode at 71.3 kHz with the modeled displacement profile in the right inset. The converted power spectral density in displacement noise unit is shown on the right *y*-axis, with the modeled thermal noise contributions. Left inset is the Lorentzian fit for the zoomed-in fundamental mode (RBW is 1 Hz), which indicates the $Q_m \approx 1,383$. b) 2D map of the transduction and optomechanical spring effect in the inertial sensor. The wavelength resolution is 2 pm on a 1552.833 nm optical resonance, at 22 µW input power. The dashed lines show the numerical modeling of the backaction transduction for different laser-cavity detunings, with the green middle line at 22 µW input and the others at 40 µW, 89 µW and 142 µW respectively. c) Measured RF shifts (Lorentzian-fitted peaks) for different laser-cavity detunings under several drive powers. Data points are illustrated, with the solid lines from the numerical modeled frequency shifts. d) Measurement setup schematic with dual-detectors for RF tracking and simultaneous optical power monitoring. Optical drive and readout are coupled through tapered fiber coupling. Devices are placed in a vacuum chamber ($\approx 10^{-6}$ torr) on a rotary stage.

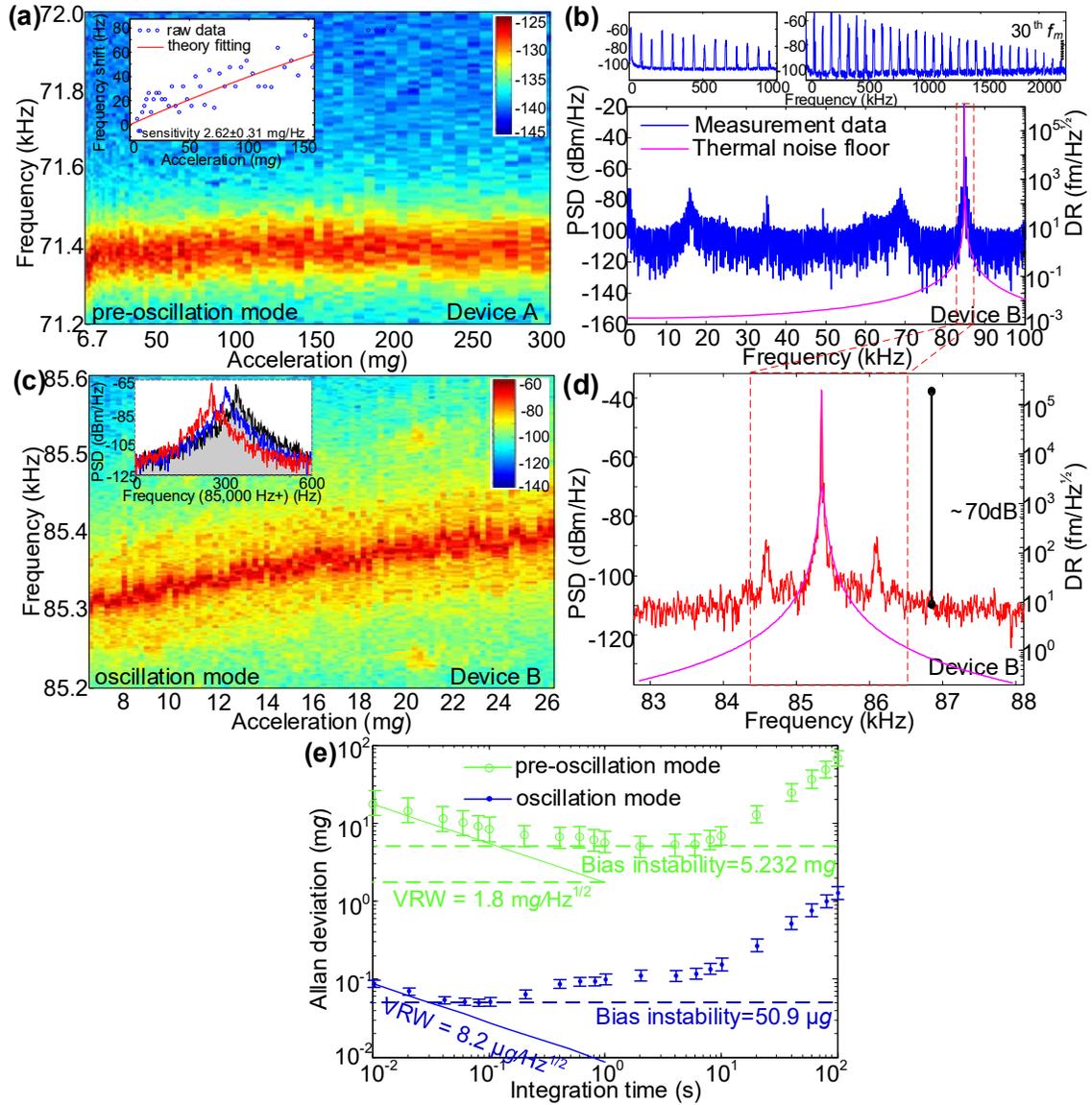

**Figure 3.** Acceleration signal detection. a) Backaction RF transduction at 40 µW drive power under different accelerations, from 6.7 m*g* to 300 m*g*. Accelerations applied on the device (Device A) pointing outwards from rotation center. Mechanical RF frequency increases by 146 Hz. Inset shows the Lorentzian fitted resonance frequency shift for increasing acceleration, with modeled solid curve. The overall fit to the detection sensitivity is at 2.62 ± 0.31 m*g*/Hz. b) Inertial sensor driven into self-sustained oscillation mode, with optical drive power exceeding the intrinsic mechanical damping, with up to 30th harmonic observed (top panel). c) Tracked mechanical power spectra of oscillation-mode acceleration sensing at ≈ 200 µW drive power. Accelerations applied are from 6.5 m*g* to 26.3 m*g* pointing outwards from rotation center, with device mounted with reduction of slot width *s*. Mechanical resonance frequency increases by ≈ 31.7 Hz. Top inset: comparison RF spectra under 6.53, 14.68 and 26.34 m*g* (from left to right) respectively. d) Zoomed-in of fundamental oscillation mode at 85.3 kHz in this device, with the sidebands arising from coupling to out-of-plane modes. e) Overall Allan deviation in terms of acceleration of the inertial sensor at pre-oscillation and oscillation modes. The obtained bias instability and velocity random walk are estimated, respectively, from the -1/2 slope and zero slope for both modes.